\documentclass[english,prd,nofootinbib]{revtex4}
\usepackage[T1]{fontenc}
\usepackage[latin1]{inputenc}
\usepackage{geometry}
\usepackage{amsmath}
\usepackage{babel}
\usepackage{graphics}
\usepackage{amssymb}
\usepackage{subfigure}

\makeatletter

\providecommand{\LyX}{L\kern-.1667em\lower.25em\hbox{Y}\kern-.125emX\@}

\makeatother
\begin{document}

\title{Soft and collinear parton radiation in heavy quark production}

\thanks{Talk presented at the American Physical Society's 2002 Meeting of The Division
of Particles and Fields (Williamsburg, VA, 2002)}

\author{\underbar{P. M. Nadolsky}\( ^{1} \), N. Kidonakis\( ^{1,2} \), F. Olness\( ^{1} \),
C.-P. Yuan\( ^{3} \) }

\affiliation{\( ^{1} \)Department of Physics, Southern Methodist University, Dallas,
TX 75275-0175\\
 \( ^{2} \)Department of Physics and Astronomy, University of Rochester,
Rochester, NY 14627-0171\\
\( ^{3} \)Department of Physics \& Astronomy, Michigan State University,
East Lansing, MI 48824}

\begin{abstract}
When the energy of the heavy quark is comparable with its mass, it is natural
to attribute this heavy quark to the hard part of the reaction. At large
energies, this approach is impractical due to large logarithms from intensive
QCD radiation affecting both inclusive and differential observables. We present
a formalism for all-order summation of such logarithms and reliable description
of heavy-quark distributions at all energies. As an illustration, we calculate
angular distributions of \( B \) mesons produced in neutral-current events
at large momentum transfers at the \( ep \) collider HERA.
\end{abstract}
\maketitle
In recent years, significant attention was dedicated to exploring properties
of heavy-flavor hadrons produced in lepton-nucleon deep inelastic scattering
(DIS). On the experimental side, the Hadron-Electron Ring Accelerator at
DESY (HERA) has generated a large amount of data on the production of charm
and bottom mesons in neutral current interactions (cf. \cite{Kroseberg:2002pb}
and references therein). At present energies (about 300 GeV in the \( ep \)
c.m. frame), a substantial charm production cross section is observed in
a wide range of Bjorken \( x \) and photon virtualities \( Q^{2} \), and
charm quarks contribute up to 30\% to the DIS structure functions. 

On the theory side, Perturbative Quantum Chromodynamics (PQCD) provides a
natural framework for the description of the heavy-flavor production. Due
to large masses \( M \) of charm and bottom quarks (\( M^{2}\gg \Lambda _{QCD}^{2} \)),
the renormalization scale can be always chosen in a region where the effective
strong coupling \( \alpha _{S} \) is small. Despite the smallness of \( \alpha _{S} \),
perturbative calculations in the presence of heavy flavors are not without
intricacies. In particular, care in the choice of a proper factorization
scheme is essential for the efficient separation of short- and long-distance
contributions to the heavy quark cross section. Recent reviews of the theoretical
status can be found in Refs.~\cite{Tung:2001mv, Hayes:1999xr, Thorne:1998xv}.
The key issue here is whether, for a given renormalization and factorization
scale \( \mu \sim Q \), the heavy quarks of the \( N \)-th flavor are treated
as \emph{partons} in the incoming proton, \emph{i.e.}, whether one calculates
the QCD beta-function using \( N \) active flavors and introduces a parton
distribution function (PDF) for the \( N \)-th flavor. A related, but separate,
issue is whether the mass of the heavy quark can be neglected in the hard
cross section and/or parton distribution function without ruining the accuracy
of the calculation. Currently, several factorization schemes \cite{Gluck:1982cp, Aivazis:1994pi, Kniehl:1995em, Buza:1998wv, Cacciari:1998it, Thorne:1998ga, Chuvakin:1999nx, Kramer:2000hn}
are available that differ in the treatment of these issues; of those, the
variable flavor number schemes reduce to the conventional massless \( \overline{MS} \)
factorization scheme in the limit \( Q^{2}\gg M^{2} \). The general proof
of the factorization in the presence of nonzero heavy quark masses was given
in Ref.~\cite{Collins:1998rz}.

In the near future, we expect the quality of the heavy flavor data to greatly
improve. By 2006, the upgraded \( ep \) collider HERA will accumulate an
integrated luminosity of \( 1\mbox {\, fb}^{-1} \), \emph{i.e.}, more than
eight times the final integrated luminosity from its previous runs. Studies
of heavy quarks in \( ep \) collisions are also envisioned at the proposed
Electron Ion Collider \cite{EICWhiteBook} and THERA \cite{Abramowicz:2001qt}.
By the end of these experiments, a large amount of information will be available
both for the inclusive observables (like the charm component of the DIS structure
functions) and differential distributions. 

In our work \cite{KNOY2002}, we address a theoretical formalism for the
description of heavy-quark differential distributions both near the heavy-quark
mass threshold (\( Q^{2}\approx M^{2}) \) and far beyond it (\( Q^{2}\gg M^{2} \)).
Our attention is primarily focused on the angular or transverse momentum
distributions of the final-state heavy mesons. The behavior of such distributions
is quite complex, since three distinct momentum scales (the heavy quark mass
\( M \), the photon virtuality \( Q \), and the transverse momentum of
the heavy quark) are involved. As a result, large logarithms of ratios of
two very different scales tend to appear in the fixed-order perturbative
cross sections in some regions of the phase space. These logarithms break
the convergence of the series in the strong coupling \( \alpha _{S} \).
To obtain stable theoretical predictions, they have to be summed through
all orders of \( \alpha _{S} \). The {}``resummation'' is realized by
the reorganization of factorized cross sections and solution of equations
for the gauge and renormalization group invariance.
\begin{figure}
{\centering \resizebox*{!}{6cm}{\includegraphics{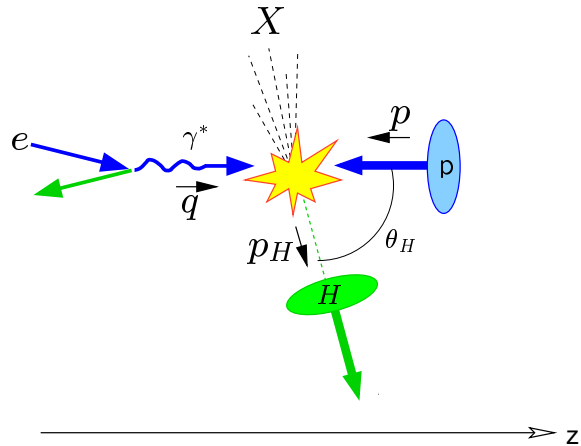}} \par}

\caption{\label{fig:GammaP}The neutral current semi-inclusive production \protect\( e+p\rightarrow e+H+X\protect \)
of heavy mesons in the \protect\( \gamma ^{*}p\protect \) c.m. reference
frame. \protect\( q^{\mu },p^{\mu },\protect \) and \protect\( p_{H}^{\mu }\protect \)
are momenta of the virtual photon, proton, and the heavy meson, respectively.
The polar angle \protect\( \theta _{H}\protect \) of the heavy meson defines
the scale \protect\( q_{T}^{2}\protect \) discussed in the text. The resummation
effects considered here are important in the limit \protect\( \theta _{H}\rightarrow 0\protect \),
\emph{i.e.}, when the final-state heavy meson closely follows the direction
of its escape in the lowest-order flavor-excitation process \protect\( \gamma ^{*}+q\rightarrow q\protect \).}
\end{figure}

The inclusive and semi-inclusive heavy-flavor cross sections may contain
large logarithms of two kinds. The logarithms \( \ln \left( Q^{2}/M^{2}\right)  \)
appear already in the inclusive observables and are resummed in the heavy-quark
PDFs. The resummation of this type constitutes the essence of the massive
variable flavor number factorization schemes, as, for instance, the ACOT
scheme \cite{Aivazis:1994pi}. Logarithms of the form \( (\alpha _{S}^{n}/q_{T}^{2})\ln ^{m}(q_{T}^{2}/Q^{2}) \),
\( 0\leq m\leq 2n-1 \) appear at \( Q^{2}\gg M^{2},q_{T}^{2} \) when calculating
angular differential distributions. Here \( q_{T}\equiv p_{T}/z \) is the
transverse momentum \( p_{T} \) of the heavy meson \( H \) rescaled by
the final-state fragmentation variable \( z\equiv (p\cdot p_{H})/(p\cdot q) \);
\( p^{\mu },q^{\mu }, \) and \( p_{H}^{\mu } \) are the momenta of the
initial-state proton, virtual photon, and heavy meson, respectively. The
limit \( q_{T}^{2}\ll Q^{2} \) is equivalent to the limit \( \theta _{H}\rightarrow 0 \),
where \( \theta _{H} \) is the polar angle of \( H \) in the \( \gamma ^{*}p \)
c.m. reference frame (cf. Fig.~\ref{fig:GammaP}). If \( M \) is neglected
(\( M^{2}\ll Q^{2} \)), \( q_{T} \) is simply related to \( \theta _{H} \):
\[
q_{T}^{2}=Q^{2}\left( \frac{1}{x}-1\right) \frac{1-\cos \theta _{H}}{1+\cos \theta _{H}}\longrightarrow Q^{2}\left( \frac{1}{x}-1\right) \left( \frac{\theta _{H}^{2}}{4}+...\right) \mbox {\, as\, }\theta _{H}\rightarrow 0.\]
The logarithms \( \ln (q_{T}^{2}/Q^{2}) \) originate from the incomplete
cancellation of soft singularities and factorization of quasi-collinear singularities
in the fixed-order cross sections.%
\footnote{A discussion of integrable singularities in the fixed-order differential
distributions for the semi-inclusive DIS charm production can be found in
Ref.~\cite{Kretzer:2001tc}.
} The formalism for their summation was proposed by Collins, Soper, and Sterman
(CSS) to describe angular correlations in \( e^{+}e^{-} \) hadroproduction
\cite{Collins:1981uk, Collins:1982va} and transverse momentum distributions
in the Drell-Yan process \cite{Collins:1985kg}. Recently this formalism
was extended to describe semi-inclusive DIS (SIDIS) production of practically
\emph{massless} hadrons \cite{Meng:1996yn, Nadolsky:1999kb, Nadolsky:2000ky}.
The resummation in massless SIDIS serves as a high-energy limit for our calculation. 

The key result of the CSS formalism is that all logarithms due to the soft
radiation in the high-energy limit can be summed into a Sudakov exponent.
This result can be summarized in the following master equation:\begin{equation}
\label{CSS}
\frac{d\sigma }{dQ^{2}\, dq_{T}^{2}}=\frac{\sigma _{0}}{s}\, \, \int \frac{d^{2}b}{(2\pi )^{2}}\, \, e^{i\vec{q}_{T}\cdot \vec{b}}\, \, \left( C^{in}\otimes f\right) \left( C^{out}\otimes d\right) e^{-S}\, \, +\sigma _{FO}-\sigma _{ASY}\, .
\end{equation}
 Here \( b \) is the impact parameter conjugate to \( q_{T} \), \( f \)
and \( d \) are parton distributions and fragmentation functions, respectively.
\( C^{in} \), \( C^{out} \) contain perturbative corrections to contributions
from the incoming and outgoing hadronic jets, respectively. The factor \( e^{-S} \)
is the Sudakov exponential, which includes an all-order sum of perturbative
logarithms \( \ln ^{m}{(q_{T}^{2}/Q^{2})} \) at \( b\lesssim 1\mbox {\, GeV}^{-1} \)
and nonperturbative contributions at \( b\gtrsim 1\mbox {\, GeV}^{-1} \).
Finally, \( \sigma _{FO} \) is the fixed-order expression for the considered
cross section, while \( \sigma _{ASY} \) (\textit{asymptotic piece}) is
the perturbative expansion of the \( b \)-space integral up to the same
order of \( \alpha _{S} \) as in \( \sigma _{FO} \). At small \( q_{T} \),
where terms \( \ln ^{m}(q_{T}^{2}/Q^{2}) \) are large, \( \sigma _{FO} \)
cancels well with \( \sigma _{ASY} \), so that the cross section~(\ref{CSS})
is approximated well by the \( b \)-space integral. At \( q_{T}\gtrsim Q \),
where the logarithms are no longer dominant, the \( b \)-space integral
cancels with \( \sigma _{ASY} \), so that the cross section (\ref{CSS})
is equal to \( \sigma _{FO} \) up to higher-order corrections. 
\vspace{0.3cm}

\begin{figure}
{\centering \resizebox*{!}{0.38\textheight}{\includegraphics{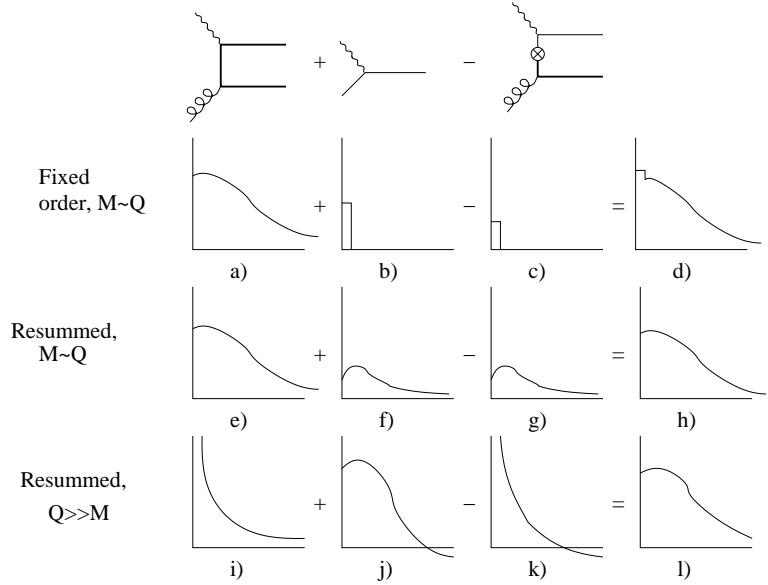}} \par}

\caption{\label{fig:Balance}Balance of various terms in the ACOT scheme and resummed
cross section. Graphs show \protect\( q_{T}\protect \) on the \protect\( x\protect \)-axis
and \protect\( d\sigma /dq_{T}^{2}\protect \) on the \protect\( y\protect \)-axis.}
\end{figure}
To extend the CSS resummation from the massless to the massive case, the
logarithms of both types \( \ln \left( Q^{2}/M^{2}\right)  \) and \( \ln \left( q_{T}^{2}/Q^{2}\right)  \)
should be resummed simultaneously. For the resummation of the logarithms
\( \ln \left( Q^{2}/M^{2}\right)  \), we use the simplified ACOT factorization
scheme \cite{Collins:1998rz, Kramer:2000hn}. With regards to the logarithms
\( \ln \left( q_{T}^{2}/Q^{2}\right)  \), we observe that the proof of the
resummation formula in the original papers \cite{Collins:1981uk, Collins:1982va, Collins:1985kg}
does not rely on the dimensional regularization for the soft and collinear
singularities appearing in individual Feynman diagrams. Although the dimensional
regularization does provide a convenient way to regulate ultraviolet singularities
in \( \alpha _{S} \) and PDFs (and hence, defines the \( \overline{MS} \)
regularization and factorization schemes), it does not have to be invoked
to regulate the singularities in the infrared region: other methods (\emph{e.g.},
regularization with nonzero quark and gluon masses) will do just fine. As
a result, the \( b \)-space integral is well-defined both for nonzero quark
masses and in the limit \( M/Q\rightarrow 0 \); hence it cannot contain
negative powers or logarithms of \( M \) with the exception of the collinear
logarithms resummed in the PDFs or fragmentation functions. Also the quark
masses regulate collinear singularities, so that at least some of the logarithms
are not dominant in the threshold region. The remaining issue is about the
accuracy of the approximation. The evolution equations used in the derivation
of the \( b \)-space integral in Refs.~\cite{Collins:1981uk,Collins:1982va}
did not keep track of the power-suppressed terms \( {\cal O}((M/Q)^{n}),\, n>0 \).
As a result, the \( b \)-space integral does not necessarily approximate
the small-\( q_{T} \) behavior of the cross section if \( M/Q \) is not
negligible. However, the full Eq.~(\ref{CSS}) does a better job and approximates
the cross section of the order \( {\cal O}(\alpha _{S}^{k}) \) up to corrections
of the order \( {\cal O}(\alpha _{S}^{k+1},M/Q) \). This happens because
terms up to the order \( {\cal O}(\alpha _{S}^{k},M/Q) \) are canceled between
the \( b \)-space integral and \( \sigma _{ASY} \); the terms \( \ln ^{m}\left( q_{T}^{2}/Q^{2}\right)  \)
that were not resummed (if any) cancel between \( \sigma _{FO} \) and \( \sigma _{ASY} \).

\begin{figure}
{\centering \subfigure{\resizebox*{0.49\textwidth}{!}{\includegraphics{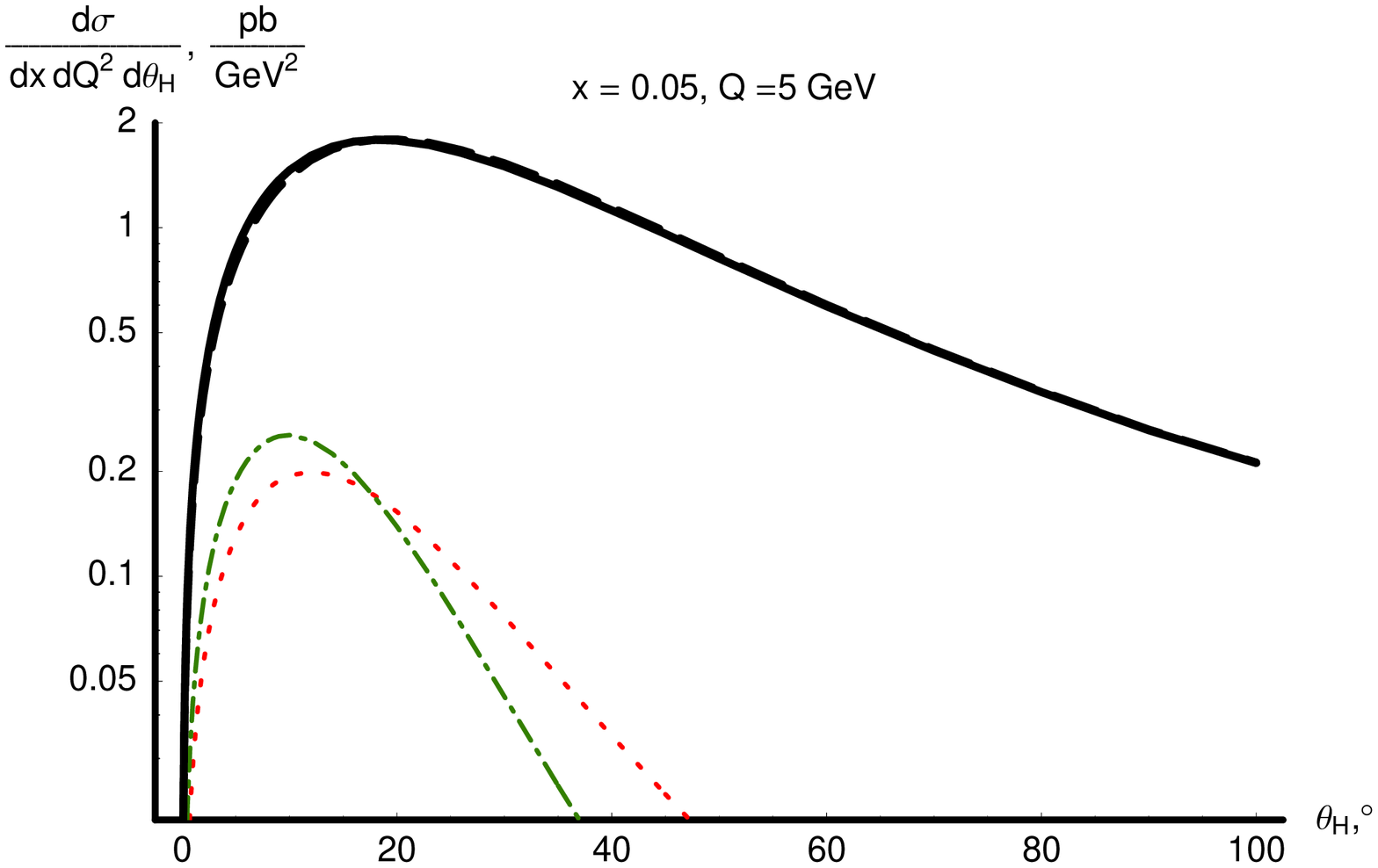}}} \subfigure{\resizebox*{0.49\textwidth}{!}{\includegraphics{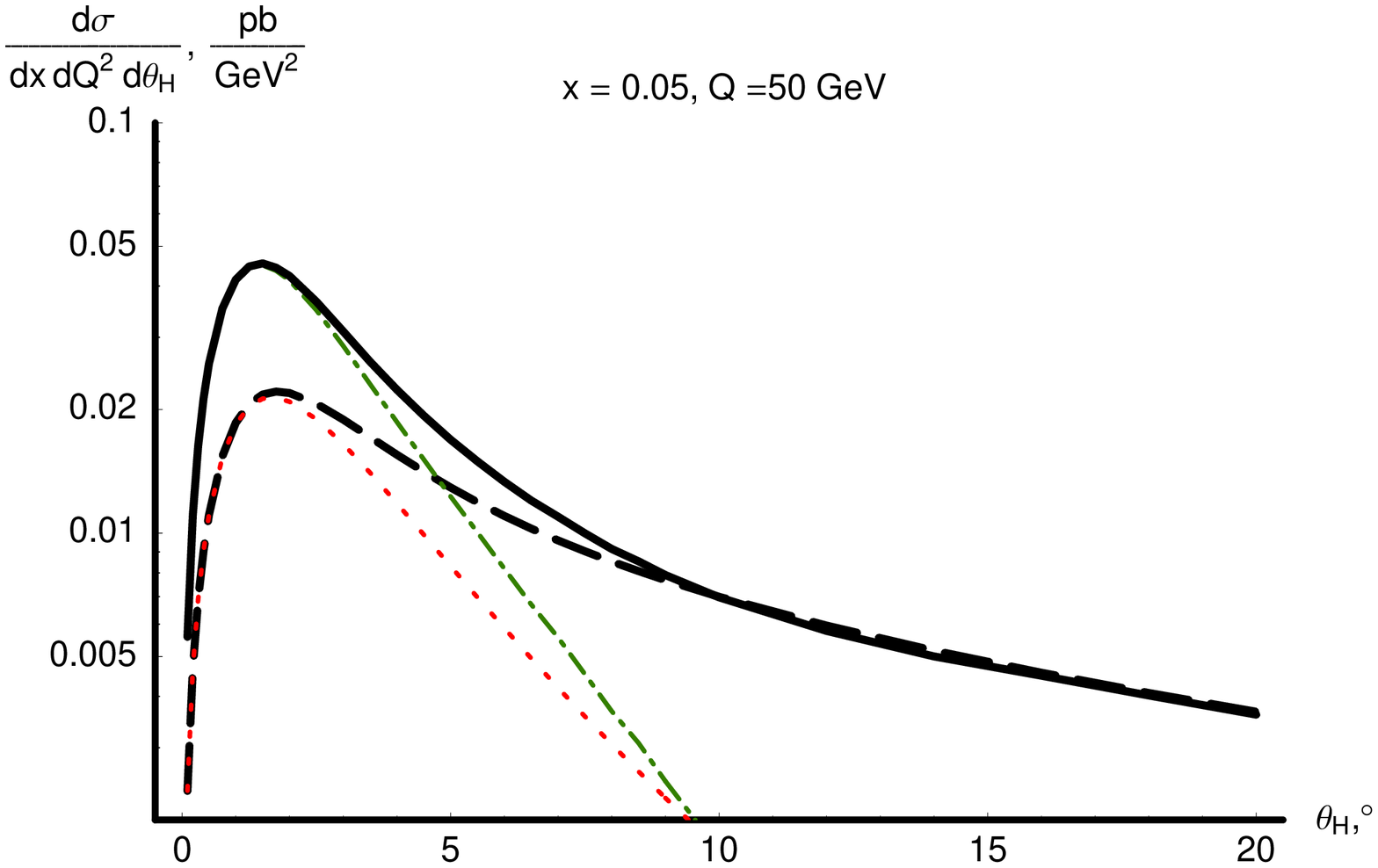}}} \par}

\caption{\label{fig:NumericalResults}The angular distributions of \protect\( B\protect \)
mesons in the \protect\( \gamma ^{*}p\protect \) c.m. frame close to the
mass threshold (\protect\( Q=5\protect \) GeV, left figure) and at \protect\( Q^{2}\gg M^{2}\protect \)
(\protect\( Q=50\protect \) GeV, right figure). In both cases, the cross
sections are calculated for \protect\( x=0.05\protect \) and \protect\( M=4.5\protect \)
GeV using the CTEQ5HQ PDFs \cite{Lai:1999wy} and Peterson fragmentation
functions \cite{Peterson:1983ak} with \protect\( \varepsilon =0.0033\protect \).
The plots show the fixed-order cross section (long-dashed line), the \protect\( b\protect \)-space
integral (dot-dashed line), the asymptotic piece (dotted line), and the full
resummed cross section (solid line). }
\end{figure}
Fig.~\ref{fig:Balance} qualitatively illustrates the balance of various
terms in the resummation formula in various regions of phase space for the
lowest-order contributions. First, consider a fixed-order calculation in
the simplified ACOT scheme near the threshold. In this region, the cross
section \( d\sigma /dq_{T}^{2} \) is well approximated by the \( {\mathcal{O}}(\alpha _{S}) \)
photon-gluon diagram (Fig.\,\ref{fig:Balance}a). To this diagram, we add
the lowest-order \( \gamma ^{*}q \) term, which resums powers of \( \ln (Q^{2}/M^{2}) \)
and contributes at \( q_{T}=0 \) (Fig.\,\ref{fig:Balance}b). We also subtract
the overlap between the two diagrams (Fig.\,\ref{fig:Balance}c), which is
approximately equal to, but not the same as, the \( \gamma ^{*}q \) contribution.
The resulting distribution (Fig.\,\ref{fig:Balance}d) is close to the fixed-order
result, but has discontinuities in the small \( q_{T} \) region. These discontinuities
are amplified when \( Q \) increases. 

In the \( q_{T} \)-resummed cross section, the fusion diagram still dominates
near the threshold, but now the resummed cross section and subtracted asymptotic
piece are smooth functions, which cancel well at all values of \( q_{T} \)
(Fig.\,\ref{fig:Balance}e-h). Hence, the distribution is physical in the
whole range of \( q_{T} \). Finally, at \( Q^{2}\gg M^{2} \) the small-\( q_{T} \)
region the \( \gamma ^{*}g \) fusion contribution is dominated by the \( 1/q_{T}^{2} \)
term (Fig.\,\ref{fig:Balance}i). Such singular terms are summed through
all orders in the \( b \)-space integral corresponding to Fig.\,\ref{fig:Balance}j
and are canceled in the fusion contribution by subtracting the asymptotic
piece (Fig.\,\ref{fig:Balance}k).

To further simplify the calculation, we use the results of Refs.\,\cite{Collins:1998rz, Kramer:2000hn},
which to neglect masses of heavy quarks entering the hard scattering subprocess
directly from the proton. This approximation, which differs from the complete
mass-dependent cross section by higher-order terms, significantly simplifies
the analysis. We therefore drop heavy-quark masses in hard scattering subdiagrams
with incoming heavy quark lines. This approximation leads to massless expressions
for the perturbative Sudakov factor and \( C \)-functions for quark-initiated
subprocesses, and mass-dependent \( C \)-functions for gluon-initiated subprocesses.

Figure \ref{fig:NumericalResults} demonstrates how various terms in Eq.~(\ref{CSS})
are balanced in an actual numerical calculation. This Figure shows the bottom
quark production cross section vs. the polar angle \( \theta _{H} \) of
the bottom quark in the \( \gamma ^{*}p \) c.m. frame for the lowest-order
processes \( \gamma ^{*}b\rightarrow b \) and \( \gamma ^{*}g\rightarrow b\bar{b} \).
At this order, the perturbative Sudakov factor is identically zero. Due to
the heavy mass of the bottom quark, the \( b \)-space integral is practically
insensitive to contributions from the nonperturbative QCD region \( b\gtrsim 1\mbox {\, GeV}^{-1} \),
so that it can be calculated without introducing 
a phenomenological parameterization for the Sudakov factor 
in the nonperturbative region. The same calculation
can be done for the charm production, but in that case the nonperturbative
Sudakov factor cannot be neglected. The left figure shows that near the heavy-quark
mass threshold (\( Q=5 \) GeV) the \( b \)-space integral cancels well
with its perturbative expansion \( \sigma _{ASY} \), so that the the full
cross section is practically indistinguishable from the fixed-order term.
On the other hand, at \( Q=50 \) GeV (right figure) the full cross section
is substantially larger than the fixed-order term at \( \theta _{H}\lesssim 10^{\circ } \)
and is dominated by the \( b \)-space integral. In this region, \( \sigma _{FO} \)
is canceled well by \( \sigma _{ASY} \). At \( \theta _{H}\gtrsim 10^{\circ } \)
the cross section agrees well with the massless result. More details can
be found in our upcoming publication.

To summarize, we presented a method for the calculation of angular distributions
in the heavy quark production both near the heavy-quark threshold and at
large momentum transfers. Our method is realized in a massive variable flavor
number factorization scheme and resums transverse momentum logarithms with
the help of the \( b \)-space resummation formalism \cite{Collins:1981uk,Collins:1982va,Collins:1985kg}.
Our approach provides an adequate tool for detailed studies of differential
distributions in a wide range of \( Q \) that will soon be available from
HERA. 

\vspace{1\baselineskip}

This work is supported by the U.S. Department of Energy, the National Science
Foundation, and the Lightner-Sams Foundation.


\end{document}